# PROGRAM SLICING TECHNIQUES AND ITS APPLICATIONS

N.Sasirekha[1], A.Edwin Robert[2] and Dr.M.Hemalatha[3]

[1]Doctoral Research Scholar, Karpagam University, Coimbatore, Tamilnadu, India
`nsasirekhaudt@gmail.com`

[2]Doctoral Research Scholar, Karpagam University, Coimbatore, Tamilnadu, India
`edrt.edwin@gmail.com`

[3]Head, Department of Software Systems, Karpagam University, Coimbatore, Tamilnadu, India
`hema.bioinf@gmail.com`

## ABSTRACT

*Program understanding is an important aspect in Software Maintenance and Reengineering. Understanding the program is related to execution behaviour and relationship of variable involved in the program. The task of finding all statements in a program that directly or indirectly influence the value for an occurrence of a variable gives the set of statements that can affect the value of a variable at some point in a program is called a program slice. Program slicing is a technique for extracting parts of computer programs by tracing the programs' control and data flow related to some data item. This technique is applicable in various areas such as debugging, program comprehension and understanding, program integration, cohesion measurement, re-engineering, maintenance, testing where it is useful to be able to focus on relevant parts of large programs. This paper focuses on the various slicing techniques (not limited to) like static slicing, quasi static slicing, dynamic slicing and conditional slicing. This paper also includes various methods in performing the slicing like forward slicing, backward slicing, syntactic slicing and semantic slicing. The slicing of a program is carried out using Java which is a object oriented programming language.*

## 1. INTRODUCTION

One of the program analysis techniques is program slicing. The main applications of program slicing include various software engineering activities such as program understanding, debugging, testing, program maintenance, complexity measurement and so on. Program slicing is a feasible method to restrict the focus of a task to specific sub-components of a program. It can also be used to extract the statements of a program that are relevant to a given computation.

The concept of a program slice was introduced by Weiser [1]. According to the original definition [2], the notion of slice was based on the deletion of statements [8]. A slice is an executable subset of program statements that preserves the original behaviour of the program with respect to a subset of variables of interest and at a given program point. Several variants of this notion have been proposed in the literature, such as dynamic slicing [5], quasi static slicing [4], simultaneous dynamic slicing [6], and conditioned slicing [7]. This paper lists out various slicing techniques and its applications.

There are various aspects to be considered in slicing a program. They are listed as follows

**Slicing variable:** Slicing variable may be based on the variables specified in the criteria (slicing point of interest) or it may be on all variables.

**Type of the result:** The result of slicing may be equivalent to the program of few set of statements from the program.

**Slicing Point:** Considering the Slicing point, programmer interest may be before or after a particular statement [4].

**Scope:** The scope of the slice may be interprocedural or it may be intraprocedural [30].

**Slicing Direction:** The expected slice from the program may be in the forward or backward direction.

**Abstraction Level:** Abstraction level is either in statement or in procedure.

**Type of Information:** The information that we obtain from the slice will be either a static or it may be a dynamic one.

**Computational Method:** The method of computing the slice is by solving the data flow equations. It may be done through graph reachability in various dependence graphs [33].

**Output format:** The format that can be obtained after slicing may seem to be in the form of code which is equivalent to source code or it may be a dependence graph or may be in the form of execution tree.

## 2. STATIC SLICING

According to Weiser [1] [2] [3], a program slice consists of the parts or components of a program that affect the values computed at some point of interest, referred to as a slicing criterion. Typically, a slicing criterion consists of a pair < S, V >, where S is the statement number and V is a variable. The components of a program which have a direct or indirect effect on the values computed at a slicing criterion < S, V > are called the program slice with respect to the slicing criterion < S, V >. The following portion of a program explains the above concept.

```
(1) DataInputStream d = new DataInputStream (System.in);
(2) terminate_var= Integer.parseInt(d.readLine());
(3) product=1;
(4) sum=0;
(5) for(counter=1; counter<=terminate_var; counter++)
    {
(6) sum=sum+counter;
(7) product=product*counter;
    }
(8) System.out.println("The Sum is : "+sum);
(9) System.out.println("The Product is : " + product);
```

Figure 1

(1) DataInputStream d = new DataInputStream(System.in);
(2) terminate_var= Integer.parseInt(d.readLine());
(4) sum=0;
(5) for(counter=1; counter<=terminate_var; counter++)
   {
(6) sum=sum+counter;
   }
(8) System.out.println("The Sum is : "+sum);

Figure 2

Figure 1 shows the portion of a program which reads any number terminate_var and calculates the sum and product of numbers from 1 to terminate_var. Figure 2 shows the slice of the program with respect to the slicing criteria (9, sum). All variables that are not relevant to the computation of the sum is sliced away. Statistically available information is used for slicing hence this type of slicing is called as static slicing.

## 2.1. Static Slicing Approach

Static slicing can be approached in terms of program reachability using Program Dependence Graph (PDG) [9][10][11]. A PDG is a directed graph with vertices corresponding to statements and control predicates, and edges corresponding to data and control dependences. The slicing criterion is identified with a vertex in the PDG, and a slice corresponds to all PDG vertices from which the vertex under consideration can be reached. The slices are computed by gathering statements and control predicates by way of a backward traversal of the program's control flow graph (CFG) or PDG, starting at the slicing criterion. Hence, these slices are referred to as backward static slices. Reps and Bricker [12] were the first to use forward static slice terminology. Informally, a forward slice consists of all statements and control predicates dependent on the slicing criterion, a statement being "dependent" on the slicing criterion if the values computed at that statement depend on the values computed at the slicing criterion, or if the values computed at the slicing criterion determine the fact if the statement under consideration is executed or not. Backward and forward slices are computed in the same way. Forward slice requires tracing dependences in the forward direction.

In other words a backward slice contains statements of a program which has some effect on the slicing criterion. It helps the developer to locate the parts of the program that contains a bug [14]. Consider the following sample portion of a program.

DataInputStream d = new DataInputStream (System.in);
terminate_var= Integer.parseInt(d.readLine());
product=1;
sum=1;
for(counter=1; counter<=terminate_var; counter++)
{
sum=sum+counter;
product=product*counter;
}
average=(sum-1)/terminate_var;
System.out.println("The Sum is : "+sum);
System.out.println("The Product is : " + product);
System.out.println("The Average is : " +average);

Figure 3

Above program (figure 3) produces a result of the variable sum with a big value. To locate a bug, backward slice conducted on the variable sum and to find out the lines that contribute to the incorrect value. Backward slice is shown in figure 4. It shows the value of sum stored as 1. Since sum is a running total it has to be initialized to zero. Therefore to correct the bug replace the assignment sum=1 with the assignment sum=0.

```
sum=1;
for(counter=1; counter<=terminate_var; counter++)
{
sum=sum+counter;
}
System.out.println("The Sum is : "+sum);
```

Figure 4

A forward slice contains the statements of the program which are affected by the slicing criterion. It is used to predict the parts that will be affected by modification in that program [14]. The forward slice for figure 3 is shown in figure 5. Forward slice on variable sum can identify the ripple effects caused due to the change in the value of the variable sum. Fixing a bug would introduce a bug in the assignment average. Hence that statement can be replaced by average=sum/terminate_var.

```
DataInputStream d = new DataInputStream (System.in);
terminate_var= Integer.parseInt(d.readLine());
product=1;sum=0;
for(counter=1; counter<=terminate_var; counter++)
{
sum=sum+counter; /* Affected statement */
product=product*counter;
}
average=(sum-1)/terminate_var; /* Affected statement */
System.out.println("The Sum is : " +sum); /* Affected statement */
System.out.println("The Product is : " + product);
System.out.println("The Average is : " +average); /*Affected statement */
```

Figure 5

From the graphical representation view (PDG), a backward slice is with respect to a slicing criterion consists of the set of nodes that directly or indirectly affect the computation of the variables in V at the node P. A forward slice is defined as the set of program statements and predicates that are affected by the computation of the value of a variable V at a program point P [13].

## 3. DYNAMIC SLICING

During program slicing, the slicing criterion contains the variables which produced an unexpected result on some input to the program [1] [3]. However, a static slice may contain statements which have no influence on the values of the variables of interest for the particular execution. During execution of a program, the value inputted may cause unexpected result. Dynamic slicing takes the input supplied to the program during execution and the slice contains only the statement that caused the failure during the specific execution of interest. Dynamic slicing uses dynamic analysis to identify all and only the statements that affect the variables of interest on the particular anomalous execution trace [15] [16]. The advantage of dynamic slicing is the run-time handling of arrays and pointer variables. Dynamic slicing will treat each element of an array individually, whereas static slicing considers each definition or use of any array element as a definition or use of the entire array [17]. Similarly, dynamic slicing distinguishes the objects that are pointed to by pointer variables during a program execution.

A dynamic slicing criterion specifies the input, and distinguishes between different occurrences of a statement in an execution. Slicing criteria consists of triple (input, occurrence of a statement, variable). The difference between static and dynamic slicing is that dynamic slicing assumes fixed input for a program, whereas static slicing does not make assumptions regarding the input. Figure 6 shows an example program to be sliced. Figure 7 shows the static slice of the program in figure 6 with the slice criterion (9, x) contains the entire program. Figure 8 shows the dynamic slice of the program in figure 6 with slice criterion (n = 2, $9^1$, x), where $9^1$ denotes the first occurrence of statement 8 in the execution of the program. For the input n = 2, the loop is executed twice. In this example (figure 6), the else branch of the if statement is omitted from the dynamic slice (figure 8) since the assignment of 18 to variable x in the first iteration of the loop is "killed" by the assignment of 17 to x in the second iteration.

```
(1) DataInputStream d = new DataInputStream (System.in);
(2) n= Integer.parseInt(d.readLine());
(3) i=1;
(4) while (i<=n)
    {
(5) if (i mod 2==0) then
(6)     x=17;
    else
(7)     x=18;
(8) i=i+1;
    }
(9) System.out.println(x);
```

Figure 6 –Sample Program to be sliced

```
DataInputStream d = new DataInputStream (System.in);
n= Integer.parseInt(d.readLine());
i=1;
while (i<=n)
{
if (i mod 2==0) then
    x=17;
else
    x=18;
i=i+1;
}
System.out.println(x);
```

Figure 7 – Static slice with criterion (9, x)

```
DataInputStream d = new DataInputStream (System.in);
n= Integer.parseInt(d.readLine());
i=1;
while (i<=n)
{
if (i mod 2==0) then
```

```
        x=17;
else
        x=18;
i=i+1;
}
System.out.println(x);
```

Figure 8 – Dynamic slice with criterion (n = 2, 91, x)

## 4. SIMULTANEOUS DYNAMIC SLICING

A different approach to the definition of a slice with respect to a set of executions of the program has been proposed by Hall [6]. This new slicing technique combines the use of a set of test cases with program slicing. The method is called simultaneous dynamic program slicing because it extends and simultaneously applies to a set of test cases the dynamic slicing technique [5] which produces executable slices that are correct on only one input.

A simultaneous dynamic slice of a program P on simultaneous dynamic slicing criterion C = ({I1, I2, … Im}, S, V) is any syntactically correct and executable program P0 that is obtained from P by deleting zero or more statements where Im refers to the Input, S is the statement in the program and V is the subset of variables in the program P. Let us consider an example program which is shown in figure 9 that finds the positive sum, negative sum, positive product, negative product. By comparing the positive, negative sum and comparing the positive and negative product, this program displays the greatest from among the two respectively.

```
(1)  int a,chk,n,i,pprod,nprod, psum,nsum,sum,prod;
(2)  DataInputStream d =new DataInputStream (System. in);
(3)  a = Integer.parseInt (d.readLine ( ));
(4)  chk = Integer.parseInt (d.readLine ( ));
(5)  n = Integer.parseInt (d.readLine ( ));
(6)  i=pprod=nprod=1;
(7)  psum=nsum=0;
(8)  while ( i<=n && a<=n) {
(9)  if (a > 0) {
(10) psum += a;
(11) pprod *= a ;}
(12) else if (a<0) {
(13) nsum -= a;
(14) nprod *=(-a);}
(15) else if (chk) {
(16) if (psum>=nsum)
(17) psum = 0;
(18) else nsum = 0;
(19) if (pprod >= nprod)
(20) pprod = 1;
(21) else nprod = 1 ;}
(22) i++;
(23) a = Integer.parseInt (d.readLine( ));
(24) if (i<=n) {
(25) sum = 0;
(26) prod = 1;}
(27) else {
(28) if (psum>=nsum)
```

(29) sum = psum;
(30) else sum = nsum;
(31) if (pprod >= nprod)
(32) prod=pprod;
(33) else prod=nprod; }
(34) System.out.println ("Sum : "+sum);
(35) System.out.println ("Product : "+prod);

Figure 9 –Sample Program to be sliced

The program slice for the Simultaneous dynamic slicing with respect to the slicing criteria C= (I1, I2, 32, {sum}) of the figure 9 is shown in the figure 10. In the slicing criteria I1 = (chk=0, n=2, a1 =0, a2=2) and I2 = (chk=1, n=2, a1 =0, a2=2).

(1)  int a,chk,n,i,pprod,nprod, psum,nsum,sum,prod;
(2)  DataInputStream d=new DataInputStream (System. in);
(3)  a = Integer.parseInt (d.readLine ( ));
(4)  chk = Integer.parseInt (d.readLine ( ));
(5)  n = Integer.parseInt (d.readLine ( ));
(6)  i = 1;
(7)  psum = nsum = 0;
(8)  while (i<=n && a<=n) {
(9)  if (a > 0) {
(10) psum += a;
(12) else if (a<0) { }
(15) else if (chk) {
(16) if (psum>=nsum)
(17) psum = 0;
(22) i++;
(23) a = Integer.parseInt (d.readLine ( ));
(24) if (i<=n) { }
(27) else {
(28) if (psum>=nsum)
(29) sum = psum;
(34) System.out.println ("Sum : "+sum);

Figure 10 –Simultaneous Dynamic Slice

A simultaneous program slice on a set of test cases is not simply given by the union of the dynamic slices on the component test cases. Indeed, simply the union of dynamic slices is unsound, in that the union does not maintain simultaneous correctness on all the inputs [6]. An iterative algorithm is presented [6] that, starting from an initial set of statements, incrementally construct the simultaneous dynamic slice, by computing the iteration a larger dynamic slice. This approach can be used in program comprehension for the isolation of the subset of the statements corresponding to particular program behaviour. It can be considered a refinement of the method proposed by Wilde et al. [36] that consider the problem of locating functionalities in code as the identification of the relation existing between the ways the user and the programmer see the program. Simultaneous dynamic slicing can be considered as a refinement of methods for localizations of functions based on test cases [36], because it takes into account the data flow of the program and then allows the reduction of the set of selected statements.

## 5. QUASI STATIC SLICING

Quasi static slicing is a hybrid of Static and Dynamic Slicing. Static slicing is examined during compile time, using no information about the input variables of the program. Dynamic slicing analyses the code by giving input to the program. It is constructed at run time with respect to a particular input. There exists trade-off between static and dynamic slicing methods. Static slicing needs more space, more resources and will perform every possible execution of the program where as Dynamic slicing needs less space and is specific to a program execution. Dynamic slices are smaller than static slice [20]. For complete program understanding one execution of the program is not enough. Hence the Quasi static slicing method was first introduced by Venkatesh [4]. In Quasi slicing the value of some variables are fixed and the program is analyzed while the value of other variables vary. The behaviour of the original program is not changed with respect to the slicing criterion. Slicing criteria includes the set of variables of interest and initial conditions and hence quasi slicing is called as Conditioned slicing [20]. This is an efficient method for program comprehension. This technique fails to prove uniform treatment of static and dynamic slicing. Moreover, there is no algorithmic description for this technique [18].

(1) DataInputStream d=new DataInputStream(System.in);
(2) n=Integer.parseInt(d.readLine());
(3) a=Integer.parseInt(d.readLine());
(4) sum=0;
(5) prod=1;
(6) if (n>0)
(7) { sum += a;
(8)    prod *=a;
(9)    a += 5; }
(10) if (n<0)
(11) { sum -=a;
(12)    prod *= a;
(13)    a -= 5; }
(14) System.out.println("Sum : "+sum);
(15) System.out.println ("Product : "+prod);

Figure 11 –Sample Program to be sliced

As an example, let us consider the portion of a program in Figure 11. The quasi static slicing with the slicing criterion [21] (C={n},1,14,{sum}) is shown in the figure 12. The slicing criterion's first argument refers to the variable whose value is to be fixed, the second argument gives the value of the variable in the first argument, the third argument shows the line number in the program to be sliced and the fourth argument indicates one of the variables in the program.

(1) DataInputStream d=new DataInputStream(System.in);
(2) n=Integer.parseInt(d.readLine());
(3) a=Integer.parseInt(d.readLine());
(4) sum=0;
(6) if (n>0)
(7) { sum += a;
(9)    a += 5; }
(14) System.out.println("Sum : "+sum);

Figure 12 –Quasi Static Slice with criterion (C={n},1,14,{sum})

## 6. AMORPHOUS SLICING

Previously discussed slicing are syntax preserving slicing where as amorphous slicing is based on preserving the semantics of the program [22][23][24]. That is, syntax preserving slicing techniques focuses on deleting statements from the program based on the slicing criterion and hence the syntax of the program statements remain the same even after slicing is performed. Where as in Amorphous slicing, is constructed using any program transformation technique which preserves the semantics of the program with respect to the slicing criterion. The slices formed are not as large as the other slicing techniques. The slice is considerably simplified form of the program with respect to the slicing criterion. Amorphous slicing assists in program comprehension, analysis and reuse. For example consider the portion of the program in figure 13, whose sliced program preserving the semantics is shown in figure 14.

for(i=0,sum=a[0],biggest=sum;i<24;sum=a[++i])
if (a[i+1] > biggest)
biggest = a[i+1];
average = sum/25;

Figure 13 –Sample Program to be sliced

The above program fragment is intended to find the biggest of an array and stores it in the variable biggest. When syntax preserving slice is applied to the program in figure 11, it gives the same set of statements. But when amorphous slicing is applied on the final value of the variable biggest, the semantics is preserved by following the constant folding technique to achieve the slicing as shown in figure 14.

for(i=1, biggest=a[0]; i<25; ++i)
if (a[i] > biggest) biggest = a[i];

Figure 14 –Amorphous slice on variable biggest

Amorphous slice on average yields the slice as follows.

Average = a[24] / 25;

The above slice indicates that the amorphous slicing is not constructed by deleting irrelevant statements but performs possible syntax transformation preserving the semantics with respect to the slicing criterion. The outcome of this slice will never be larger than the original program to be sliced.

## 7. APPLICATIONS

The applications of program slicing techniques has now ramified into a powerful set of tools for use in such diverse applications as program understanding, program verification, automated computation of several software engineering metrics, software maintenance and testing, functional cohesion, dead code elimination, reverse engineering, parallelization of sequential programs, software portability, reusable component generation, compiler optimization, program integration, showing differences between programs, software quality assurance, software fault-injection [26][27][30][31][32]. Program slicing is very useful in extracting the business rules from legacy systems.

## 7.1 Debugging

Programmers slice the programs mentally when they attempt to debug them [1][3][34]. In order to help this mental debugging became the motivation behind the development of an automated technique that would assist debugging activities by removing the fault to a number of lines or statements in the program. That is the code that is not relevant to a set of variables the fault originated from, and therefore could not have caused the fault, would be omitted from the program. This allows the debugger to concentrate on the lines or statements that are significant to the fault, and thus help in pinpointing it. Dynamic slicing in particular is appropriate when applying program slicing to debugging, as it produces smaller slices and makes available the inputs that caused the fault [35]. Debugging process includes various stages like

**(1) Finding errors:** Execute the program by a given input. If the result is not coincident with the expected result, it indicates that there are errors in the program. Suppose their may be error when we input the value of a variable V at statement i.

**(2) Localizing errors:** Localizing errors is a trivial and time-consuming work in debugging. It is difficult to find errors quickly and precisely without tools. In order to find errors, insert breakpoints into the program. The purpose is to narrow the errors to the program section before the breakpoint. The i in a slicing criterion <i, V> is a breakpoint in nature and V is the variable set interesting. But program slice only includes the statements that influence the V at i, not all the statements before i. Thus, the error codes are localized to a less program section. If the slice is executable [15], the errors are to be localized by executing the slice, not the whole program repeatedly.

**(3) Modifying errors**: Modify the program according to the specification given by the user or it may be from the document.

**(4) Analyzing ripple effort [25]:** Program is a system with interactions among statements. The modification of an error (statement) might cause other unexpected errors. Thus when a statement is modified we must check the statements affected by the modification and modify them if necessary. The process to analyze the affected statements is called ripple analysis [25]. Such statements can be obtained from the slicing.

## 7.2 Cohesion Measurement

Cohesion for the programs is measured as an application of syntax-preserving static program slicing. A cohesive program is one in which the modularization of the program is performed correctly. A cohesive function or procedure should perform the tasks that are related to each other.

Slicing application in cohesion measurement plays a vital role in object oriented programming language. Cohesion theme in the object-oriented programming is based on encapsulation. A well encapsulated code contains all the necessary data and member function associated with that object. Program slice captures a thread through a program, which is concerned with the computation of some variable. The idea of functional cohesion is that a member function should perform related tasks. Several slices from a function, each for a different variable can be sliced. Each of these slices may have a lot of code in common, and it can be justified thinking that the variables were related in some way. Decision on the function's tasks can be captured by the computation it performs on these variables, and therefore if can be concluded that the function's tasks were strongly related and that the function was thus highly cohesive.

Let us consider an example for the actions which are related. A function that performs division of two numbers calculates the quotient and remainder. Here the results are related to one another.

Consider another function which returns the largest of two numbers and their product. This function is less cohesive. The motivation for assessing the cohesiveness of a program or a part of it rests upon observations and claims that highly cohesive programs are easier to maintain, modify and reuse. Cohesion was a product of the effort to define principles for programming, turning the activity from a craft into an engineering discipline.

A slice captures a thread through a program, which is concerned with the computation of some variables. The idea of functional cohesion is that a function should perform related tasks. If several slices from a function are considered, each for a different variable, and the slices have a lot of code in common, then it can be justified that the variables were related in some way and decide that the function's tasks are captured by the computation it performs on these variables. Therefore if can be concluded that the function's tasks were strongly related and that the function was thus highly cohesive.

### 7.3 Comprehension

In the Software development life cycle, the maintenance phase starts with program comprehension. This is absolutely acceptable in the case of legacy systems, where documentation may be sparse and the original developers may no longer be available. Conditioned slicing can help with the comprehension phase of maintenance [37]. Using constrained slicing comprehension can be done [38]. Both the types of slicing share the property that a slice is constructed with respect to a condition in addition to the traditional static slicing criterion. The condition can be used to identify the cases of interest in slicing.

### 7.4 Maintenance and Re-engineering

Software maintenance is often followed by re-engineering effort, whereby a system is manipulated to improve it. Canfora at al. [39] introduced conditioned slicing in order to allow conditions to be used to extract code based on conditions. The idea is to isolate desired functionality so that it can be recovered and reused. This application of conditioned slicing to reuse is further developed by Cimitile, De Lucia and Munro [37] as part of the reverse engineering project. Gallagher and Lyle [40] showed how a decomposition slice can be built for each variable in a program, the decomposition slices are used to divide up the software with a view to limiting the impact of software changes during maintenance activities. The decomposition slice has a complement that contains the parts of the software which can be altered without affecting computation of the variable for which the decomposition slice is constructed. This allows the programmer to altering the software, without causing unintentional changes due to the ripple effects of the changes.

### 7.5 Testing

Slicing helps to decompose programs which in testing, it makes test work faster and more efficient. Slicing based on particular slice criteria. Through this the inter-related modules can be identified, which then can be tested separately with out disturbing the rest of the program. Because program slicing helps in understanding programs by dividing it into slices, the task of testing can be allocated to a various testers. Each tester can test the slice in the program domain [22].

## 8. RELATED WORK

Program slicing was originally introduced in 1984 by Mark Weiser. He defined a slicing criterion as any subset of program variables at a statement. The program slice consists of those statements that may affect the values of the criterion variables, including whether or not the statement executes. Since then, many researchers have extended it in many directions and for all International Journal of Software Engineering & Applications (IJSEA), Vol.2, No.3, July 2011programming paradigms. The huge amount of program slicing-based techniques has lead to the publication of different surveys [29] [41] [42] [43] [45] [44] that elucidate the differences between them. However, each survey presents the techniques from a different perspective. For instance, [29] [41] [42] [43] [35] mainly focus on the advances and applications of program slicing-based techniques, but in contrast [41] focuses on the implementation and comparisons of their empirical results. Harman et al. in 1996 tries to compare and classify the various techniques in order to predict their applications. This paper combines the above authors' views and extends the idea in java based programs.

## 9. CONCLUSION

This paper discussed about various types of slicing techniques like the static slicing, dynamic slicing, quasi static slicing, amorphous slicing, forward and backward slicing which comes under the broad category of syntactic slicing and semantic slicing. Each category of slicing is explained in detail with separate program portion which was developed in java language. The application of slicing in various areas like debugging, cohesion measurement, comprehension, maintenance and re-engineering and testing are highlighted in this paper.

## 10. FUTURE WORK

Extended analysis of slicing in case of Concurrent Object oriented programs, Distributed object oriented programs has to be focused. Slicing techniques can be performed through the graph based approaches like Control Flow Graph (CGF), Program Dependence Graph (PDG), Object oriented Program Dependency Graph (OPDG), Dynamic Dependence Graph (DDG), Dynamic Object oriented Dependence Graph (DODG), Object oriented Dependency Graph(ODG), Class Hierarchy Sub graph (CHG), Control Dependence Sub graph (CDS) and Data Dependence Sub graph(DDS). Algorithmic approach to slicing can also be focused.

## REFERENCES

[1]    M. Weiser,(1982), "Programmers use slices when debugging", *Communications of the ACM*, 25(7):446 – 452.

[2]    M. Weiser, (1979), "Program slices: Formal, psychological and practical investigations of an automatic program abstraction method", Ph.D. Thesis, University of Michigan, Ann Arbor, ML.

[3]    M. Weiser, (1984), "Program slicing", *IEEE Trans. Software Engineering* , Vol. 10, pp 352-357.

[4]    G.A. Venkatesh, "The semantic approach to program slicing", *ACM SIGPLAN* Notices, vol. 26, no. 6, 1991, pp. 107-119.

[5]    B. Korel and J. Laski, (1990), "Dynamic slicing of computer programs", *The Journal of Systems and Software*, vol. 13, no. 3, pp. 187-195.

[6]    R.J. Hall,( 1995), "Automatic extraction of executable program subsets by simultaneous program slicing", *Journal of Automated Software Engineering*, vol. 2, no. 1, pp. 33-53.

[7]    G. Canfora, A. Cimitile, and A. De Lucia,( 1998), "Conditioned program slicing", *Information and Software Technology*, vol. 40, no. 11/12, pp. 595-607.

[8]    M. Harman and S. Danicic, (1997), "Amorphous program slicing", Proceedings of 5th International Workshop on Program Comprehension, Dearborn, Michigan, U.S.A., IEEE CS Press, pp. 70-79.

[9]    K.J. Ottenstein and L.M. Ottenstein, (1984), "The program dependence graph in a software development environment", In Proceedings of the ACM SIGSOFT/SIGPLAN Software Engineering Symposium on Practical Software Development Environments, pages 177–184, SIGPLAN Notices 19(5).61

**Authors**

**N.Sasirekha**, completed MCA, M.Phil. in Computer Science. She is a doctoral research scholar in Computer Science at the Karpagam University, Coimbatore, Tamilnadu, India. She is currently working as Assistant Professor in PG Department of Computer Applications at Vidyasagar College of Arts and Science, Udumalpet, Tamilnadu. She has presented eleven papers in various National Conferences, Seminar and one paper in International Conference. Her area of research is Software Engineering.

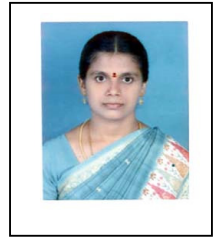

**A.EdwinRobert**, MCA., M.Phil., in Computer Science. He is a Doctoral research scholar in Computer Science at the Karpagam University, Coimbatore, Tamilnadu,India. He is currently working as Assistant Professor in Software Systems Department at Karpagam University, Coimbatore, TamilNadu. He had five years of teaching experience. He has presented a paper in National Conference. His area of research is Software Engineering.

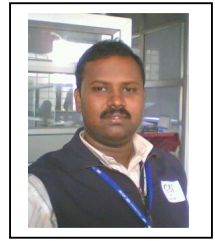

**Dr. M. Hemaltha,** completed MCA MPhil., PhD in Computer Science and Currently working as a Asst Professor and Head , Dept of Software Systems in Karpagam University. She has ten years of experience in teaching and published Twenty Seven papers in International Journals and also presented Seventy papers in various National Conferences and one paper in International Conference. Her area of research are Data mining, Software Engineering, Bioinformatics and Neural Networks. She is also a reviewer in several National and International Journals.

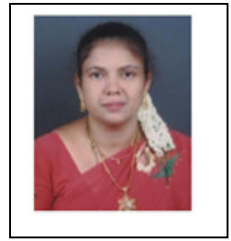